# The Space Infrared Interferometric Telescope (SPIRIT): High-resolution imaging and spectroscopy in the far-infrared

David Leisawitz[a], Charles Baker[a], Amy Barger[b], Dominic Benford[a], Andrew Blain[c], Rob Boyle[a], Richard Broderick[a], Jason Budinoff[a], John Carpenter[c], Richard Caverly[a], Phil Chen[a], Steve Cooley[a], Christine Cottingham[d], Julie Crooke[a], Dave DiPietro[a], Mike DiPirro[a], Michael Femiano[a], Art Ferrer[a], Jacqueline Fischer[e], Jonathan P. Gardner[a], Lou Hallock[a], Kenny Harris[a], Kate Hartman[a], Martin Harwit[f], Lynne Hillenbrand[c], Tupper Hyde[a], Drew Jones[a], Jim Kellogg[a], Alan Kogut[a], Marc Kuchner[a], Bill Lawson[a], Javier Lecha[a], Maria Lecha[a], Amy Mainzer[g], Jim Mannion[a], Anthony Martino[a], Paul Mason[a], John Mather[a], Gibran McDonald[a], Rick Mills[a], Lee Mundy[h], Stan Ollendorf[a], Joe Pellicciotti[a], Dave Quinn[a], Kirk Rhee[a], Stephen Rinehart[a], Tim Sauerwine[a], Robert Silverberg[a], Terry Smith[a], Gordon Stacey[f], H. Philip Stahl[i], Johannes Staguhn[j], Steve Tompkins[a], June Tveekrem[a], Sheila Wall[a], and Mark Wilson[a]

[a] NASA's Goddard Space Flight Center, Greenbelt, MD
[b] Department of Astronomy, University of Wisconsin, Madison, Wisconsin, USA
[c] California Institute of Technology, Pasadena, CA, USA
[d] Lockheed Martin Technical Operations, Bethesda, Maryland, USA
[e] Naval Research Laboratory, Washington, DC, USA
[f] Department of Astronomy, Cornell University, Ithaca, USA
[g] Jet Propulsion Laboratory, California Institute of Technology, Pasadena, CA
[h] Astronomy Department, University of Maryland, College Park, Maryland, USA
[i] NASA's Marshall Space Flight Center, Huntsville, Alabama, USA
[j] SSAI, Lanham, Maryland, USA

## ABSTRACT

We report results of a recently-completed pre-Formulation Phase study of SPIRIT, a candidate NASA Origins Probe mission. SPIRIT is a spatial and spectral interferometer with an operating wavelength range 25 - 400 μm. SPIRIT will provide sub-arcsecond resolution images and spectra with resolution R = 3000 in a 1 arcmin field of view to accomplish three primary scientific objectives: (1) Learn how planetary systems form from protostellar disks, and how they acquire their inhomogeneous composition; (2) characterize the family of extrasolar planetary systems by imaging the structure in debris disks to understand how and where planets of different types form; and (3) learn how high-redshift galaxies formed and merged to form the present-day population of galaxies. Observations with SPIRIT will be complementary to those of the James Webb Space Telescope and the ground-based Atacama Large Millimeter Array. All three observatories could be operational contemporaneously.

## 1. INTRODUCTION

In 2004 NASA solicited Origins Probe Mission Concept Study proposals and selected the Space Infrared Interferometric Telescope (SPIRIT) and eight other concepts for study. Previously SPIRIT had been recommended in the US *Community Plan for Far-Infrared/Submillimeter Space Astronomy*[1] as a pathfinder to the Submillimeter Probe of the Evolution of Cosmic Structure (SPECS),[2] which had been accepted into the NASA astrophysics roadmap following recommendations of the Decadal Report *Astronomy and Astrophysics in the New Millennium*.[3] While SPECS is conceived as a 1-kilometer maximum baseline far-IR interferometer requiring multiple spacecraft flying in a tethered formation,[2,4,5] SPIRIT was intended by the infrared astronomical community as a less ambitious yet extremely capable mission, and a natural step in the direction of SPECS.

The IR astrophysics community in the US recommended SPIRIT knowing that far-IR observatories providing angular resolution orders of magnitude better than that of the current and near-future telescopes will be needed to answer a number of compelling scientific questions: How do planetary systems form from the disks of material commonly found around young stars? Why do some planets end up being hospitable to life as we know it, while others do not? How did galaxies form and evolve, occasionally colliding, erupting with newborn stars, and emerging from their dusty cocoons with freshly-forged chemical elements? The magnificent data returned from the 85-cm diameter Spitzer Space Telescope hint at the progress that will be made when sharper infrared images can be obtained. While Spitzer's far-IR angular resolution is comparable to that available to Galileo at visible wavelengths four centuries ago, SPIRIT will provide far-IR images a hundred times sharper than those of the Multiband Imaging Photometer[6] on Spitzer (MIPS), and SPECS, perhaps a decade later, will further improve the image quality by another order of magnitude.

To maximize the scientific return from the SPIRIT mission within a nominal Origins Probe cost cap we studied three alternative engineering designs, evaluated their measurement capabilities, and estimated their costs. A Design Reference Mission representing a wide range of possible research objectives was used to assess the scientific value of each design concept. After assimilating lessons from two preliminary design studies and establishing science priorities we chose a final set of measurement requirements and developed "Design C." We refined the instrument design, derived spacecraft bus requirements, developed plans for the Integration and Test program and technology development, and estimated the mission cost to be approximately $800M for development, in-orbit checkout, and three years of scientific operation. A standard parametric space hardware cost model provided independent validation of the estimated cost. Design C was reviewed by an external Advisory Review Panel.

SPIRIT Design C will enable us to accomplish the scientific objectives described in Section 2 of this paper. These are central goals in the science plan for NASA's Astrophysics Division and they relate directly to the Agency's mandate to "conduct advanced telescope searches for Earth-like planets and habitable environments around other stars."

Only a space-based far-IR observatory can make the observations needed to achieve the scientific objectives listed above and described in Section 2.1. Protostars, proto-planetary disks, planetary debris disks, and young galaxies radiate most of their energy at mid and far-IR wavelengths. Visible and near-IR light is emitted less strongly, and severely attenuated by foreground dust. This obscuring material is ubiquitous in the places of interest because all stars and planets are born in dense, dust-laden molecular clouds. Far-IR light from these stellar and planetary nurseries can readily penetrate the dust and reach our telescopes. While longer-wavelength (millimeter and sub-millimeter) light also reaches Earth and, indeed, penetrates our atmosphere where it can be measured with ground-based telescopes, only the far-IR spectrum harbors the information necessary to answer the questions posed above. The Earth's atmosphere absorbs far-IR light, and high-altitude ambient-temperature telescopes are limited in sensitivity because the photons emitted by the telescope swamp those of interest from the sky. Thus SPIRIT, like the predecessor missions IRAS (Infrared Astronomical Satellite),[7] COBE (Cosmic Background Explorer),[8] ISO (Infrared Space Observatory),[9] and Spitzer,[10] will be a cryogenically cooled space-based observatory.

SPIRIT has two 1 m diameter light collecting telescopes and a Michelson beam combining instrument. The telescopes can be separated by distances ranging up to 36 m, and the optical delay line in the beam combiner can be scanned to provide, simultaneously, sub-arcsecond angular resolution images and $R = \lambda/\Delta\lambda = 3000$ spectral resolution. Cryo-cooled optics and sensitive detectors enable SPIRIT's sensitivity to be limited only by photon noise from the sky.

Additional features of the SPIRIT C design are presented in this paper, which is organized as follows. The SPIRIT science goals and the measurements required to achieve them are detailed in Section 2. Section 3 describes the engineering implications of the measurement requirements and most of the important aspects of the C design. Information about SPIRIT's unique capabilities relative to those of other current and planned facilities is given in Section 4, which is followed by a short summary in Section 5.

# 2. SCIENCE WITH SPIRIT

To facilitate mission planning we reviewed many potential scientific objectives for a far-IR spatial and spectral interferometer, prioritized the objectives, and developed a Design Reference Mission (DRM). The DRM outlines the measurement capabilities (number of targets, wavelength range, field of view, angular and spectral resolution, sensitivity, and dynamic range) required to achieve all of the considered potential objectives. This enables us to measure the extent to which alternative mission design concepts satisfy the measurement requirements. Only the measurement requirements corresponding to the three scientific objectives classified as "primary objectives" were allowed to drive the mission design.

In Section 2.1 we describe the three primary objectives of the SPIRIT mission. Additional possible applications for the observatory are listed in Section 2.2. In Section 4 the measurement capabilities of SPIRIT are compared with those of current and near-future far-IR observatories, and with two facilities whose observations will be complementary to those of SPIRIT: the James Webb Space Telescope (JWST)[11] and the Atacama Large Millimeter Array (ALMA).[12]

## 2.1. Primary scientific objectives

We prioritized science objectives according to their perceived importance to advancing astrophysical knowledge, their relevance to the higher-level science objectives of NASA's Astrophysics Division, the plausibility that a cost-constrained far-IR mission like SPIRIT could make the required measurements, and the extent to which *only* SPIRIT could make the necessary measurements. Three objectives were considered to satisfy these criteria, and were adopted as "primary scientific objectives" for the mission: with SPIRIT we will: (1) learn how planetary systems form from protostellar disks, and how they acquire their inhomogeneous composition; (2) characterize the family of extrasolar planetary systems by imaging the structure in debris disks to understand how and where planets of different types form; and (3) learn how high-redshift galaxies formed and merged to form the present-day population of galaxies.

Below we summarize each of the primary objectives and explain how SPIRIT will be used to achieve them.

### 2.1.1. Goal 1: Learn how planetary systems form

The greatest obstacle to our understanding how planetary systems form is the unavailability of an observing tool that can decisively constrain theoretical models. The early phase of planet formation takes place behind a veil of dust, rendering protoplanetary disks inaccessible to visible wavelength telescopes. Further, until now, the mid and far-infrared light from protostars and protoplanetary disks, which is their dominant emission, has only been seen as an unresolved blur or, worse, as the blended emission of many objects in a single telescope resolution element. Only spatially-resolved far-IR observations of these objects have the power to break model degeneracy and teach us how planetary systems form. Sub-arcsecond angular resolution will be needed to resolve the structures of interest (Figure 1). To study the early gas-rich phase, when giant planets form, we will need to resolve objects in regions like ρ Oph and Taurus, at 140 pc. The gas-poor terrestrial planet formation phase can be observed in the Tuc or TW Hya associations whose distances are ~50 pc. The Spitzer Space Telescope and the planned observatories Herschel[13] and SOFIA (the Stratospheric Observatory for Infrared Astronomy)[14] lack the necessary resolution by more than an order of magnitude in the far-IR.

To gain astrophysical insight into the planet formation process, in addition to resolving protoplanetary disks spatially, we will have to exploit information available in the spectral domain (Figure 1). The far-IR continuum emission is dominated by thermal emission from dust. A spatial map of the far-IR spectral energy distribution will enable future observers to measure the three-dimensional distribution of solid state material and probe the dust temperature locally. By measuring the gas contents of planet forming disks of various ages it will be possible to constrain the timescale for gas giant planet formation and the migration of planetary bodies of all sizes. Direct observations of $H_2$ and HD in the readily excited 28 μm and 112 μm rotational lines will lessen our dependence on surrogate gas tracers, such as CO, which can be photodissociated or frozen onto grain surfaces. Numerous far-IR lines from hydrides, such as CH and OH, will be measured with Herschel and mapped with SPIRIT, as will the strong [C I], [O I] and [C II] fine structure lines at 370, 146, 63 and 158 μm. When coupled with models,[15-17] far-IR spectral line observations will give us tremendous insight into the chemistry and physical conditions in young planet forming disks. The C/O ratio, derivable from these observations, is thought to affect the composition, surface chemistry, and perhaps the habitability of planets.[18,19]

Because of its biological significance, water will be one of the most interesting molecules to observe. Most of the volatiles found on the terrestrial planet surfaces are thought to be delivered by impacts of small $H_2O$ rich bodies from the outer solar system. ALMA will probably not be able to detect water vapor through the atmosphere, or will detect it rather noisily. SPIRIT will access the rich far-IR spectrum of the $H_2O$ molecule (Figure 1c, in red), map the water distribution in young protoplanetary disks to study the formation of the water reservoir, and search for evaporating water in extrasolar comet trails.[20,21] The interferometer will make complementary observations of frozen water by mapping the $H_2O$ ice features at 44 and 63 μm.[22]

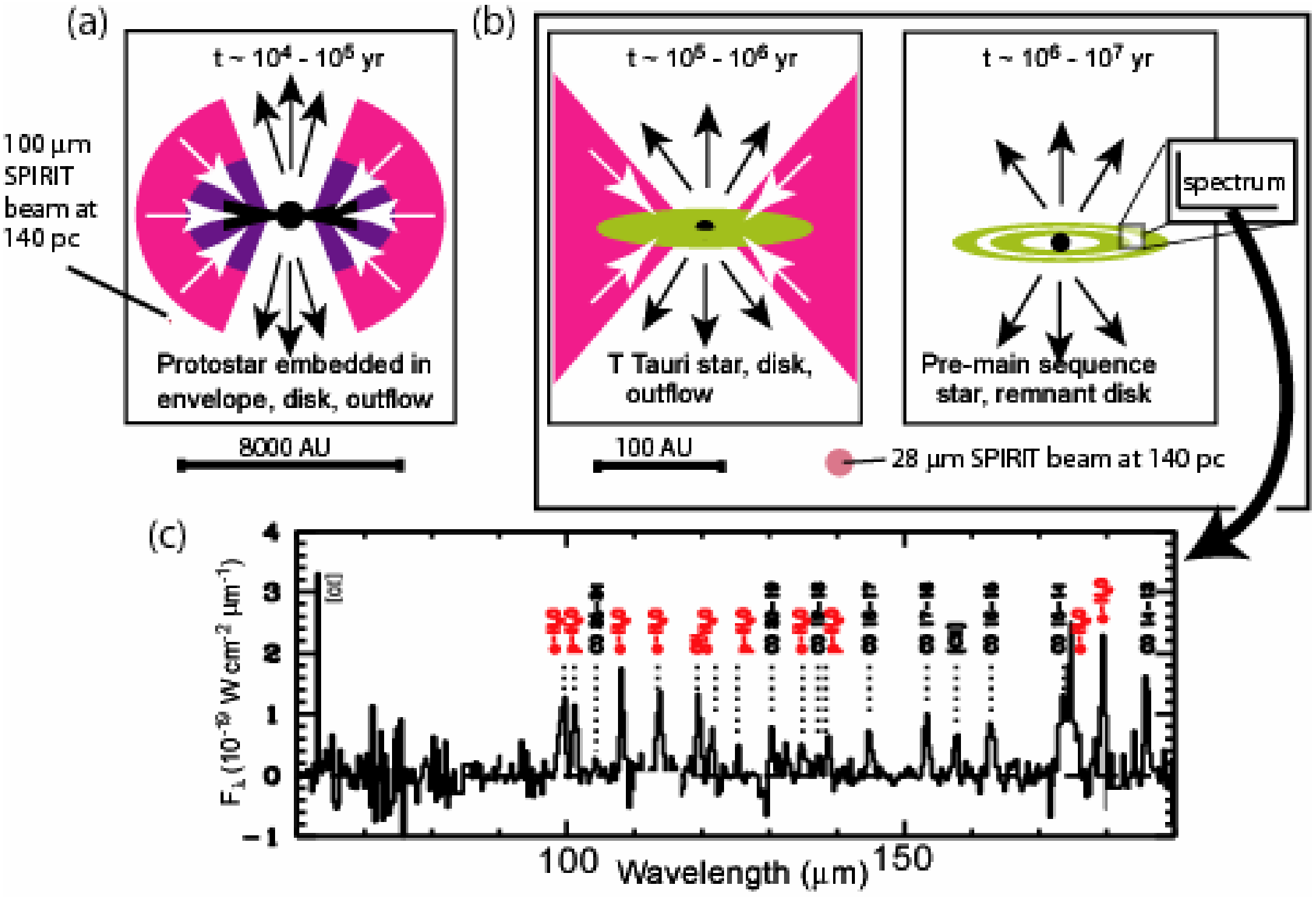

**Figure 1** – SPIRIT, which naturally produces spatial-spectral "data cubes," will (a) image protostellar disks in their line and continuum emission, and (b) map $H_2$, HD, $H_2O$ (gas and solid states), simple hydride, and C and O fine structure line emission in protoplanetary disks, providing a line spectrum as rich as the one shown in panel (c) at each of many spatial locations. The continuum-subtracted spectrum in (c) is based on ISO observations of a Class 0 protostar.[23]

As illustrated in Figure 1, SPIRIT will provide the sub-arcsecond angular resolution needed to resolve protostellar disks and planetary systems during their formative stages. SPIRIT will be used to image hundreds of these objects in the nearby regions of low-mass star formation Taurus, ρ Oph, and Perseus. Such observations will permit us to calibrate the effects of variable viewing geometry and, effectively, to finely time-resolve the development stages of a protoplanetary disk. SPIRIT will provide emission line maps in the most easily excited $H_2$ rotational transition at 28 μm at nine times better angular resolution than JWST, complementing JWST observations of the 17 and 28 μm lines, and obviating our dependence on traditional, notoriously unreliable surrogate tracers of gas density. To summarize, SPIRIT will be able to map the distribution of dust continuum and gas spectral line emission in protostars and protoplanetary disks to fill an information void, constrain theoretical models, and greatly improve our understanding of the planet formation process.

#### 2.1.2. Goal 2: Observe debris disks to characterize extrasolar planetary systems

Spatially resolved far-IR images of debris disks can reveal hidden planets in a manner complementary to the motion-dependent (astrometric or Doppler) techniques. The orbits followed by dust grains in developing and established planetary systems (i.e., debris disks) are perturbed gravitationally by the planets. Orbital resonances can produce dust concentrations,[24, 25] which have been observed at visible to millimeter wavelengths.[26 - 29] More massive planets leave more prominent signatures in debris disks, but the Earth is trailed by a dust concentration,[30] so the possibility exists that even so-called "terrestrial" planets can be found in images of debris disks. The locations, masses and orbits of unseen planets can be deduced from the shapes and temporal variations in the dust debris disks, just as new Saturnian moons have been found after ring gaps and features divulged their hiding places. This interplanetary dust glows most brightly in the spectral range ~20 – 300 μm, where main sequence stars are faint.

Spitzer is revolutionizing debris disk research and teaching us that dust replenishment in evolved planetary systems may be sporadic.[31] As noted in the previous section, however, a lack of imaging capability has permitted degeneracy in possible interpretations to go unchecked. Although the nearest debris disks are only a few parsecs away and resolved by Spitzer/MIPS[6] (Figure 2a), an important objective is to understand our own solar system in the context of a representative sample of exoplanetary systems. At a minimum, it will be necessary to detect 1 AU structures in debris disks out to a distance of ~10 pc, implying an angular resolution requirement of 0.1 arcsec. At 25 μm, JWST will miss that target by an order of magnitude, but SPIRIT will be able to image 19 known luminous debris disks[32] and many more faint disks (Figure 2b).

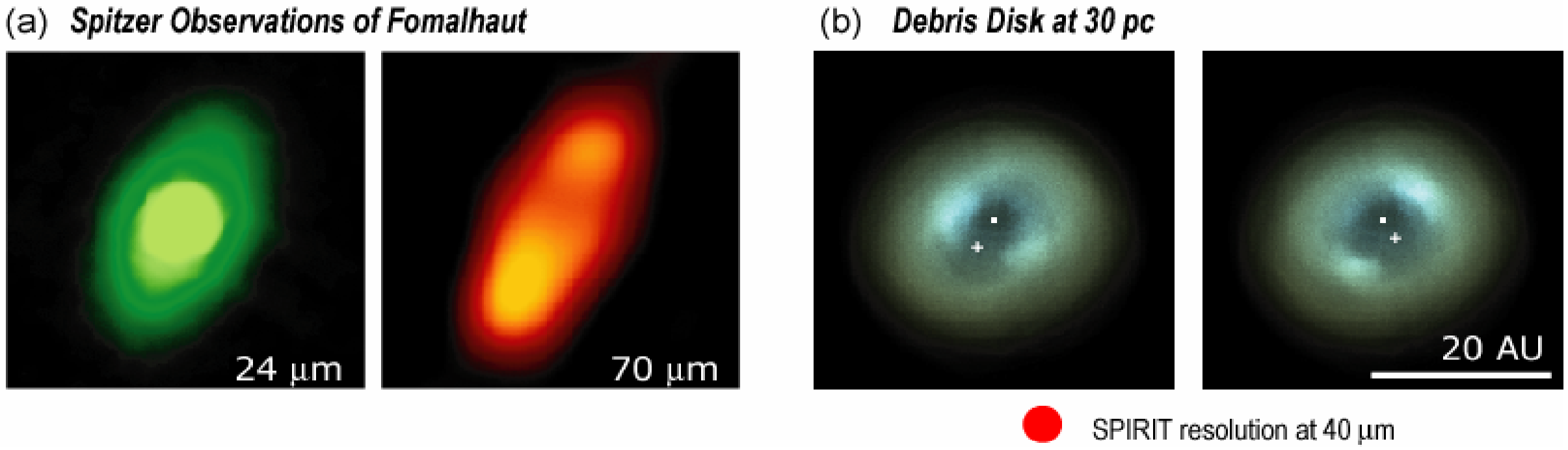

**Figure 2** – Spitzer resolves four nearby debris disks, including Fomalhaut, shown here (a) at 24 and 70 μm.[27] With angular resolution a hundred-fold better than that of Spitzer, SPIRIT will provide clear images of a large statistical sample of debris disks, enabling discoveries of new planets and a great improvement in our understanding of the factors that influence the evolution of planetary systems. The model images in (b), based on Eps Eri but scaled to 30 pc, show the predicted far-IR emission at 40, 60, and 100 μm color-coded as blue, green, and red, respectively. The dust-trapping planet (+) is shown at two orbital phases, and the resonantly trapped dust grains can be seen to have moved.

Vital to planning the Terrestrial Planet Finder Interferometer (TPF-I)/Darwin mission,[33 - 35] SPIRIT will have the sensitivity to detect a 3 "zodi" disk (1 zodi = $10^{-6} L_{Sun}$) or map the distribution of dust in a resolved 100 zodi disk at 10 pc in a 1-week observation period. Later, SPECS will have the angular resolution and sensitivity to see the signatures of a large number of low-mass planets in debris disks out to ~30 pc.[2]

#### 2.1.3. Goal 3: Learn how galaxies form and evolve

Mergers and interactions between protogalaxies played a major role in the morphological development of galaxies and their star forming histories, but compelling cosmological questions will go unanswered until better measurement capabilities exist. What is the merger history of galaxies, and which processes shaped the diverse galaxy types found in the universe today? Is the light emitted by merging galaxies predominantly due to star formation or matter accretion onto supermassive black holes? What are the dynamics of merging systems? What is the dust and chemical enrichment history of the universe, and what role did this enrichment play in the formation of galaxies and the radiation they emit?

Information uniquely available in the far-IR is critical to our understanding of the galaxy-building process. The Cosmic Background Explorer (COBE) showed that roughly half of the observed flux from galaxies resides in the far-IR.[36] The origin of this radiated energy needs to be known if we are to account for both the gravitational and nuclear energy released since the first stars began to shine and the first massive black holes collapsed. This requires investigations at comparable sensitivity and angular resolution at both far-IR/submillimeter and optical/near-IR wavelengths.

While extinction-corrected optical/near-IR data suggest a star formation rate per unit comoving volume rising moderately between z ~ 6 and z ~1, then dropping sharply from z ~1 to z ~ 0, [37 - 39] past, current, and near future far-IR telescopes lack the combination of angular resolution, sensitivity, and spectroscopic capability needed to test this conclusion by determining if the individual emitting objects are powered predominantly by star formation or by mass accretion onto a nuclear black hole. SPIRIT will have the powerful suite of measurement capabilities needed to address this problem. Many young galaxies and protogalaxies have IR-dominated spectra, and some have no optical counterparts whatsoever.[40] Galaxy mergers compress and shock the interstellar medium, and trigger bursts of star formation,[41] but inherent to this mechanism is its occurrence in dense, dust-laden clouds, which prevent the escape of stellar UV/visible photons. Most of the short-wavelength radiation is absorbed locally, where it warms interstellar dust to a temperature at which it glows most brightly in the far-IR. Similarly, galactic nuclear sources are often obscured from view at short wavelengths, but they heat the surrounding interstellar medium and produce strong far-IR emission. Further, the IR spectral fingerprints of stellar nurseries and Active Galactic Nuclei (AGN) are readily distinguishable. For example, the line ratios from a triad of neon fine-structure lines differentiates between these emission mechanisms.[42]

Far-IR and submillimeter observations are also essential to determine the merger history of galaxies. Larger galaxies are thought to have formed through successive merging of increasingly massive galaxies, and aggregates of blue stars that could represent these merging subunits are seen in the deep fields surveyed by the Hubble Space Telescope. However, puzzles remain. For example, the existence of high-redshift QSOs, the observed chemical evolution of elliptical galaxies, and the high mass measurements of submillimeter galaxies[43] all point to the conclusion that large structures were already in place by the time the universe was only several hundred Myr old. Key to understanding the merger history is to distinguish between small fragment merger systems and objects that have already formed into large disks and spheroids. In order to see the merger processes, micro-Jansky continuum sensitivity and arcsecond angular resolution are required at the far-IR wavelengths where merging galaxies are especially luminous, both in the local universe and at high redshifts.

To understand the merger process, we also need the ability to measure the kinematics of merging objects and their post-merger products in order to learn their internal dynamics, find merging substructures, and map the gas flows that fuel star formation and AGN. Many spectral lines, including the [CII] 158 μm line, the strongest line emitted by the Milky Way, will be detectable over a wide redshift range to provide kinematic information and determine spectroscopic redshifts. In order to measure the kinematics of merging galaxies, a spectral resolution of ~3000 is required at wavelengths $\lambda > 158$ μm. The Heterodyne Instrument for the Far-Infrared (HIFI) instrument on Herschel[44] will spectrally resolve the 158 μm [C II] line and provide valuable reconnaissance of nearby galaxies, but Herschel HIFI will lack the sensitivity and angular resolution needed to conduct detailed studies of the merger kinematics in distant extragalactic sources. The Spitzer IRS instrument operates in the 5.3 – 38 μm wavelength range and provides spectral resolution $\lambda/\Delta\lambda$ up to 600, enabling, for example, complementary observations of the mid-IR Ne fine structure lines and redshift determinations,[45] but Spitzer also lacks the angular and spectral resolving power needed to solve this problem.

Heavy elements and dust in the interstellar medium boost the cooling efficiency and should promote star formation. The very young universe was chemically pristine, yet evidence for metal-rich objects at redshift as high as z = 6 already exists from the Sloan Digital Sky Survey. [46] This survey probes the most extreme galaxies, but it provides no insights into the metal and dust enrichment of the vast majority of galaxies, the likely progenitors of galaxies observed today. Detailed insight into the enrichment history of the universe will come from far-IR measurements of thermal radiation from newly formed dust, spectral features of polycyclic aromatic hydrocarbons (PAHs), atomic and ionic fine-structure transitions, and rotational or vibrational transitions from small hydrides (e.g., OH), $H_2O$, CO, and $H_2$.

To break theoretical degeneracies and definitively resolve the open questions posed above, we will need integral field spectroscopy with spectral resolution R ~ 3000, sub-arcsecond angular resolution, wavelength coverage from ~25 μm to ~100(1+z) = 400 μm (for z = 3), a field of view ~1 arcminute in diameter, and sensitivity to emission at the micro-Jansky per beam level. ALMA will extend these capabilities around the SED peak for galaxies at z > 3.

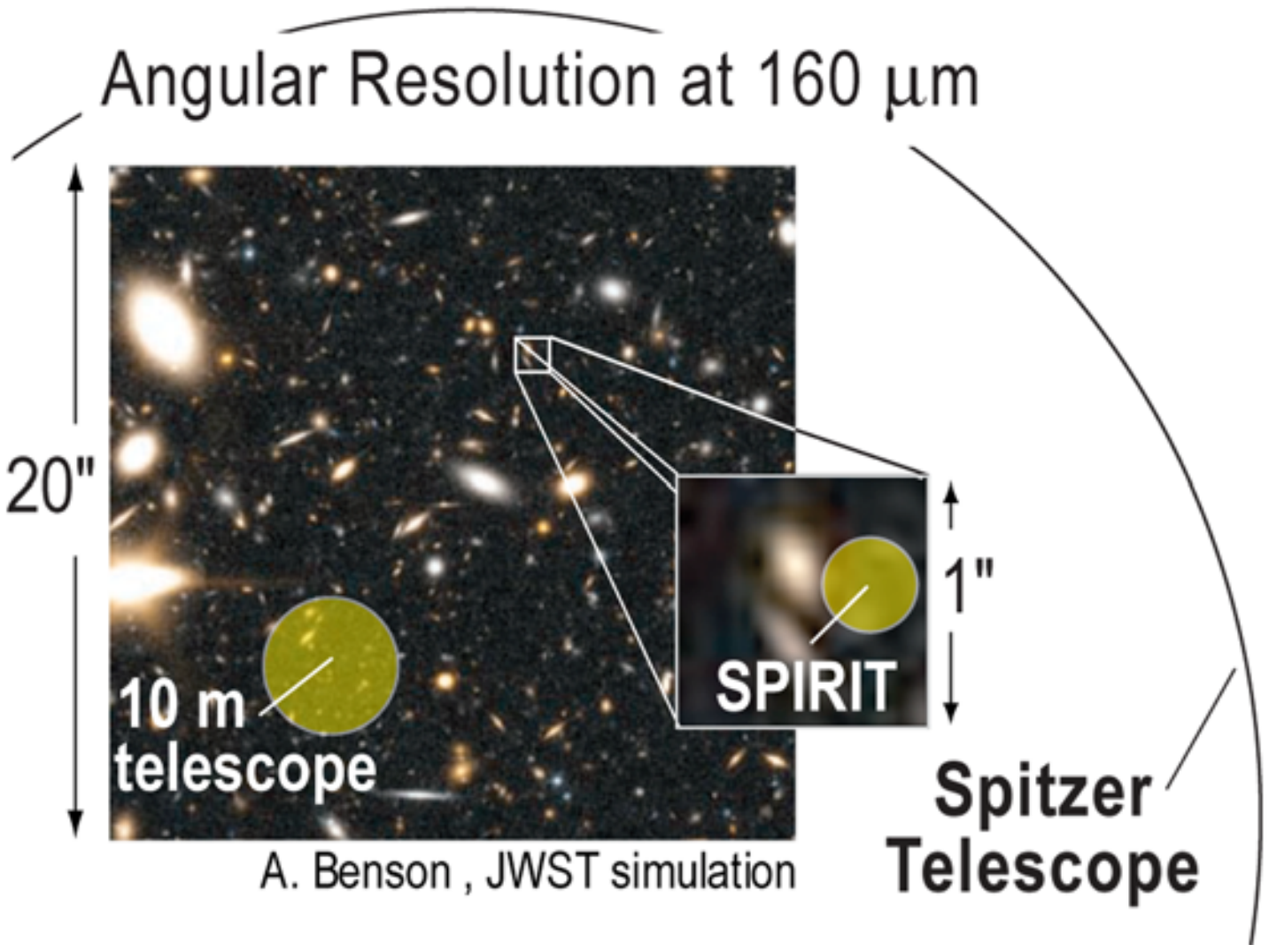

**Figure 3** –A simulated JWST deep field observation is used to illustrate SPIRIT's ability to distinguish the emissions of individual high-z objects. For comparison, the Spitzer Space Telescope resolution at 160 μm is coarser than the entire 20 arcsecond field shown, and even a 10 m diameter Single Aperture Far-IR (SAFIR) telescope would see multiple objects per beam at this wavelength.

Unlike Spitzer/MIPS,[6] Herschel,[13] SOFIA,[14] and even the 10 m diameter SAFIR Telescope,[47] SPIRIT will be able to observe galaxies in the far-infrared without the adverse effects of source confusion (Figure 3), and it will be able to measure the information-rich far-IR spectra of these objects (Figure 4). JWST will observe starlight from the high-redshift universe, while SPIRIT and ALMA will provide complementary information essential to our understanding of galaxy formation and the related processes of star formation and heavy element synthesis.

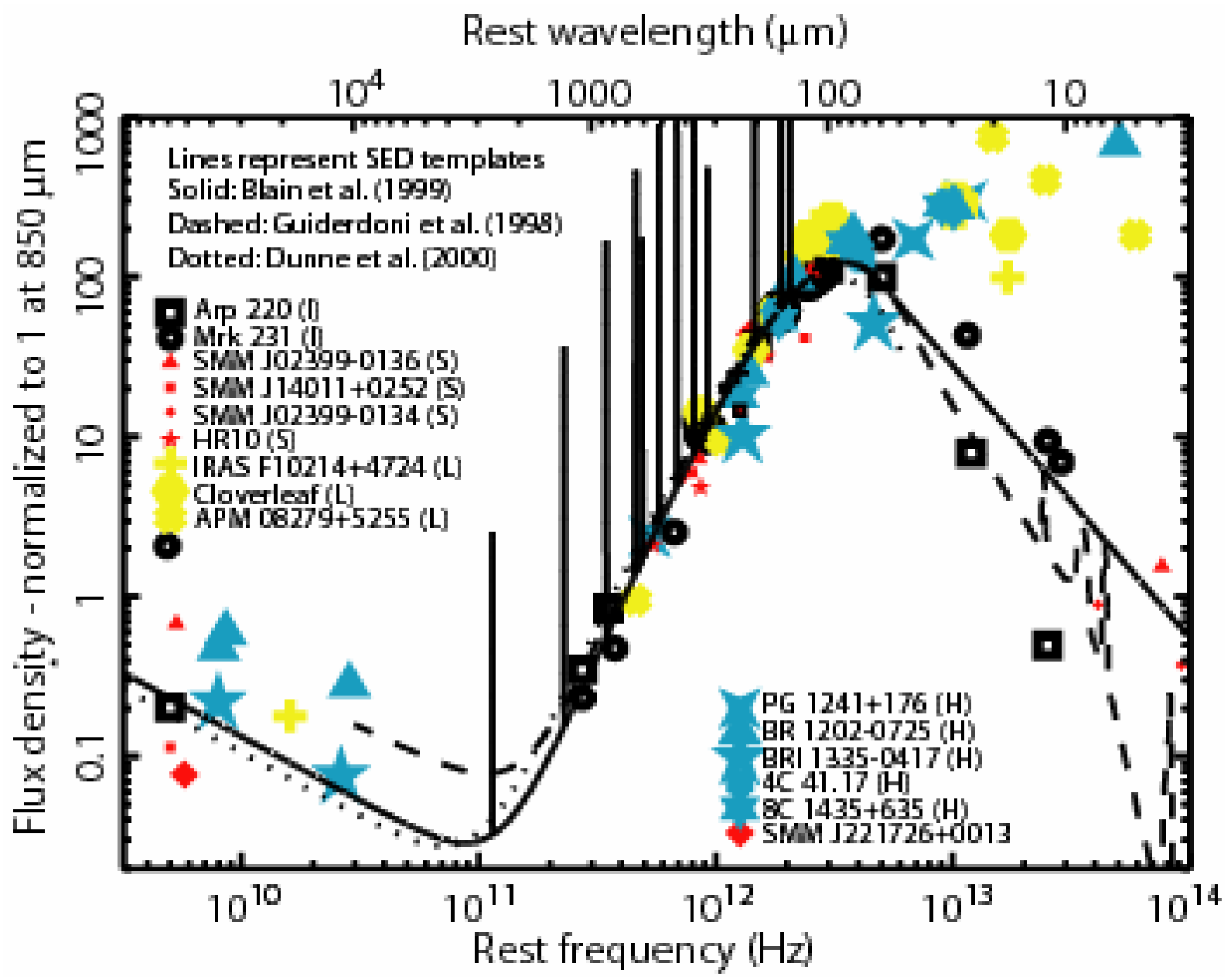

**Figure 4** –The far-IR spectrum of a galaxy[40] contains a large fraction of the bolometric luminosity and many strong emission lines, which can be used to calculate physical conditions in the interstellar medium, estimate rates of star formation, and measure a galaxy's redshift. SPIRIT will be able to measure the line intensities and the far-IR spectral energy distributions of galaxies out to high redshifts to probe the epoch of most-intense star formation (z ~ 2 - 3), distinguish between nuclear and non-nuclear emission, and determine the relationship between galaxy mergers and star formation activity.

### 2.2. Additional possible scientific applications for SPIRIT

We envision that Guest Observers will use SPIRIT to address a wide range of interesting astrophysical questions. As part of our science goal prioritization process, we considered many possible applications for a far-IR interferometer in addition to the three primary objectives described above. The following set of applications illustrates the potential of a SPIRIT mission to solve a variety of astrophysical problems: (1) Test the unification model for AGN by measuring the physical conditions in the tori surrounding the nuclei of nearby galaxies; (2) measure the spectra of extrasolar gas giant planets to determine their effective temperatures and assess their chemical diversity; (3) map nearby spiral galaxies in fine structure and molecular lines to better understand the role of large-scale gas compression in the process of star formation; (4) measure the temperature and mass structures in "starless cores" and infrared-dark clouds to study the earliest stages of star formation; and (5) follow up Spitzer observations of Kuiper Belt and other cold objects in the solar system to mine the fossil record of planetary system formation.

## 3. SPIRIT MISSION CONCEPT

Table 1 summarizes the measurement requirements derived from the scientific objectives described in Section 2 and gives the basic parameters of the SPIRIT C Design. SPIRIT will cover the wavelength range 25 to 400 μm; the angular resolution will be 0.3 arcseconds at 100 μm and scale in proportion to the wavelength; the spectral resolution will be 3000; the instantaneous field of view (FOV) will be 1 arcminute; and the field of regard will be 40 degrees wide and centered on the ecliptic plane (at any given time during the year, SPIRIT will be able to observe a point in the sky within 20 degrees of the anti-Sun direction). In addition, we required that the micro-Jansky continuum emission and the important interstellar cooling and diagnostic spectral lines from galaxies at redshift $z \sim 3$ be detectable in a “deep field” exposure lasting approximately two weeks.

A typical SPIRIT observation will take about 1 day and yield, after ground data processing, a “data cube” with two high-resolution spatial dimensions and a third, spectral dimension. These integral field spectroscopic data can be explored either as a sequence of images, each of which shows the appearance of the target field at a discrete wavelength, or as a set of spectra, one for each position in the FOV. At each field position, or for a set of discrete objects in the field, SPIRIT will measure a chemical fingerprint and indicators of physical conditions, such as gas and dust temperature and density, enabling astronomers to achieve the objectives outlined above.

**Table 1**. SPIRIT Measurement Requirements and Design Parameters

| | |
|---|---|
| Wavelength range | 25 to 400 μm |
| Instantaneous field of view | 1 arcmin |
| Angular resolution | 0.3 (λ/100 μm) arcsec |
| Spectral resolution (λ/Δλ) | 3000 over entire wavelength range |
| Point source sensitivity | $3\times10^{-18}$ W $m^{-2}$ spectral line, 5σ in 1 hr<br>~1 μJy continuum in “deep field” |
| Typical time per target field | 29 hrs |
| Field of regard | 40° band centered on ecliptic plane |
| Telescopes | 2 afocal Cassegrain |
| Primary mirror diameter | 1 m |
| Telescope temperature | 4 K |
| Interferometric baselines | 6 to 36 m |
| Orbit | Sun-Earth L2 |
| Mission life, on station | 3 years (propellant for 5) |

A single scientific instrument will provide the required measurement capabilities. SPIRIT has two 1-m diameter light-collecting telescopes and a central beam-combiner attached to a rigid structure. The telescopes are mounted on trolleys, which move along rails to provide interferometric baselines ranging in length from 6 m to 36 m. The observatory rotates during an observation with the rotation axis pointing toward the target field. The resulting spatial frequency “u-v plane” coverage can be tailored according to the expected spatial brightness structure in the scene and can be dense, so SPIRIT will produce very good images. For each baseline observed, an optical delay line is scanned, yielding a set of white light interferograms, one for each pixel in SPIRIT's four pairs of detector arrays. SPIRIT is a "double Fourier" interferometer[48 - 51] because spectral information is available in these interferograms; the delay line is scanned through the distance required to (a) equalize path lengths through the two arms of the interferometer for all angles in a 1 arcmin wide FOV, and (b) provide spectral resolution R = 3000.

While all of the essential elements of mission design were addressed during the SPIRIT study, including, for example, the attitude control system, power system, and flight software, below we describe only the novel aspects of the optical, cryo-thermal, detector, metrology, and mechanism subsystems, and the SPIRIT operations concept.

### 3.1. Optical system design

SPIRIT's light-collecting telescopes sample two sections of an incident wavefront and direct compressed, collimated beams into a beam-combining instrument. An off-axis telescope design (Figure 5) was chosen to avoid wavelength-dependent diffraction effects which would complicate calibration and compromise the sensitivity to fringe visibility. The telescopes compress the beam by a factor of 10 to minimize the total optical surface area, and therefore the thermal mass that must be cryo-cooled to meet the sensitivity requirement. Compression at the collector telescopes by more than the optimal ratio magnifies off-axis field angles and increases the size of the mirror required to intercept the beam at the beam combiner. Compression by less than the optimal ratio requires a large tertiary mirror at the collector telescope, which is both harder to cool and harder to steer to maintain alignment with the beam combiner optics.

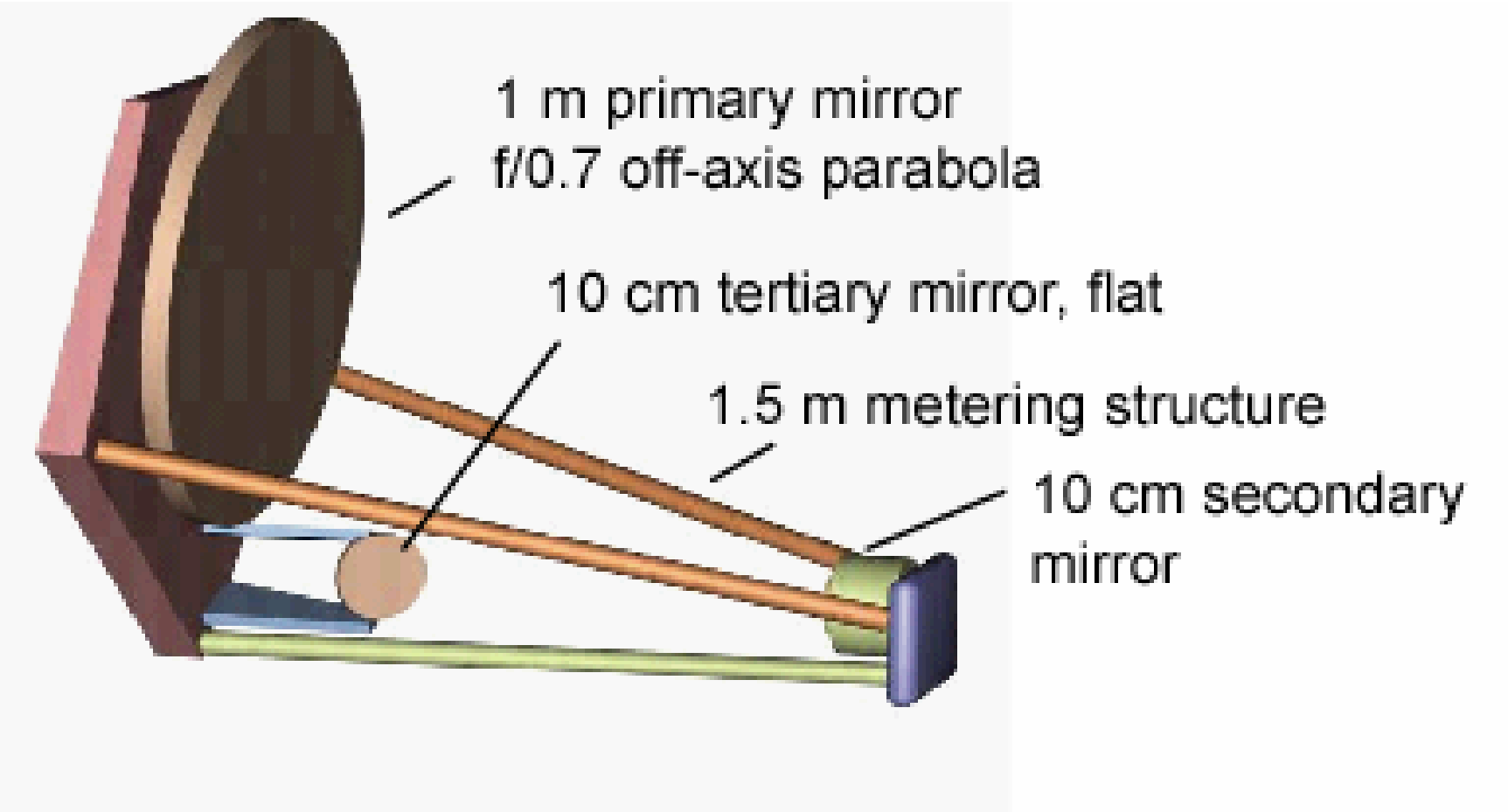

**Figure 5** – The SPIRIT light collectors are off-axis, afocal Cassegrain telescopes.

SPIRIT has a Michelson (pupil plane) beam combiner. This type of instrument permits the use of a smaller number of detector pixels and has relaxed alignment tolerances compared with the alternative Fizeau design, and the Michelson technique naturally enables Fourier Transform Spectroscopy. To develop the technique of wide field-of-view double Fourier interferometry we built the Wide-field Imaging Interferometry Testbed (WIIT),[49] which is essentially a scale model of SPIRIT. Our experience with WIIT[50] informed our design choices, and our recently-developed optical system model of the testbed[51] will ultimately serve as the basis for a high-fidelity performance model of SPIRIT.

As illustrated in Figure 6, light incident upon the SPIRIT beam combiner is further compressed with off-axis optics, recollimated, split into four wavelength bands with metal mesh dichroics to optimize optical system and detector performance, directed through optical delay lines up to the point of beam combination at a metal mesh 50/50 beamsplitter, and focused with off-axis reflectors onto detector arrays. An opto-mechanism in one collimated beam ("PCM" in Figure 6) provides pathlength correction with control feedback from an internal metrology system. The other collimated beam passes through a scanning optical delay line. In each wavelength band, both Michelson output ports are used, so there is no wasted light and the design is robust against a detector failure. The optical design is symmetric; the light collected at one telescope is reflected by mirrors and reflected or transmitted by dichroics identical in design to the optical elements encountered by light from the other telescope. This symmetric arrangement balances the throughput, simplifies calibration, and minimizes instrumental visibility loss. In the 25 – 50 μm wavelength band the beam arriving at each light collecting telescope traverses 12 optical components (mirror surfaces, dichroic, and beamsplitter) and we estimate the total optical system transmittance to be 78%. The beam is reflected by more optics in the longer wavelength bands, reducing the optical system transmittance to 72%, 67%, and 63% in the 50 – 100 μm, 100 – 200 μm, and 200 – 400 μm bands, respectively.

The optical elements have no extraordinary fabrication or material requirements. Surface roughness and mid-spatial frequency tolerances are such that some of the customary polishing steps may be avoided. Aluminum, silicon carbide, composite, and beryllium are all acceptable materials, and aluminum was adopted in the C design. An important developing capability is the capacity to produce "designer" dichroics and beamsplitters with wavelength-dependent transmission and reflection curves approaching the ideal rectangular shapes in the SPIRIT spectral range.

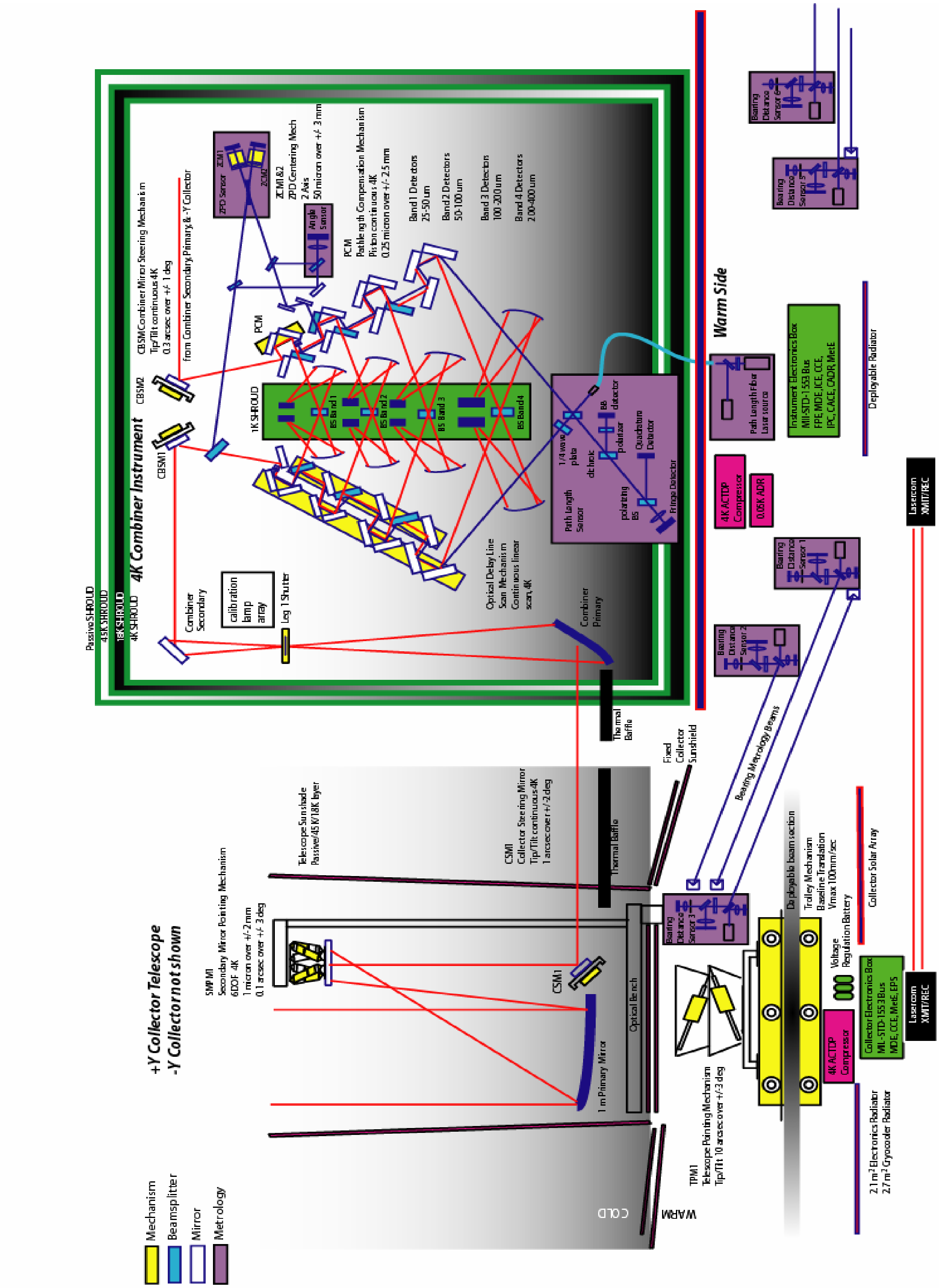

**Figure 6** – Schematic diagram showing major SPIRIT instrument components. Only one light-collecting telescope is shown.

### 3.2. Thermal design and cryocoolers

SPIRIT's thermal performance is crucial to its ability to satisfy the sensitivity requirements. Instrumental noise, either stray thermal radiation or detector noise, must not exceed the photon noise from the sky. Thus, the optics must be cooled to 4 K. We developed a thermal model and gave careful thought to thermal system design and stray light baffling. We modeled the radiative, conductive, and mechanism-generated parasitic heat loads from all sources, and used conservative estimates and large margins for the relatively poorly known terms in the thermal model. Many features of the SPIRIT design, such as low-conductivity structures and tailored, actively cooled multi-layer sun shades and shields, were optimized to satisfy thermal system performance requirements. The result is a system whose cooling requirements can be met with cryocoolers only modestly enhanced relative to those currently under development for JWST under NASA's Advanced Cryocooler Technology Development Program (ACTDP). A subscale engineering testbed has been designed, built and tested to validate the SPIRIT and related thermal models. Inner shield temperature data from the initial test match the predicted temperatures to within 0.1 K out of 20 K. The experiment also demonstrated the viability of double-aluminized Kapton as shield material, even at the lowest shield temperature, providing a lightweight solution. A space validation experiment is under consideration for the New Millennium mission ST-9.[52]

Cryocoolers have many advantages over cryostats. Elimination of a mission lifetime-limiting expendable in favor of a device whose design lifetime can easily surpass the desired mission duration is an obvious plus. In a mission like SPIRIT, where the scientific performance, especially angular resolution, is limited by the fairing dimensions in available, affordable launch vehicles, the vastly reduced volume of a cryocooler relative to a cryostat of equivalent cooling capacity is a tremendous advantage. Cryocoolers are also much less massive than cryostats and their associated support structures, a factor which can also lead to lower launch cost. Dewars also introduce unique Integration & Test (I&T) and launch operations challenges which can be avoided if cryocoolers are used.

Separate ACTDP cryocoolers are provided for each of SPIRIT's telescopes and for the beam combiner, and each of these major observatory components has a redundant cryocooler to reduce risk. The beam combiner also contains a continuous adiabatic demagnetization refrigerator (CADR) with a 4K "warm" interface, which provides detector cooling to 30 mK. A CADR has been demonstrated to operate down to 35 mK, and 20 mK is considered to be attainable in the near future.

Instead of shading the entire boom structure, we elected to shade only the components that must be actively cooled: the telescopes and the beam combiner instrument module. A boom-sized sun shade would have been at least 40 m long, challenging to deploy, and it would have unnecessarily consumed precious volume in the launch shroud. The SPIRIT multi-layer shades (Figure 7) are sized and shaped to protect cold components, including inner shade layers, from sunlight (at angles up to 20 degrees from the anti-Sun direction) and thermal radiation sources on the observatory. The telescope shades move along the boom with the telescopes. The beam combiner shade shields the instrument from the warm spacecraft. Additional shields and baffles enclose the telescopes and contribute to passive cooling and the elimination of stray thermal radiation.

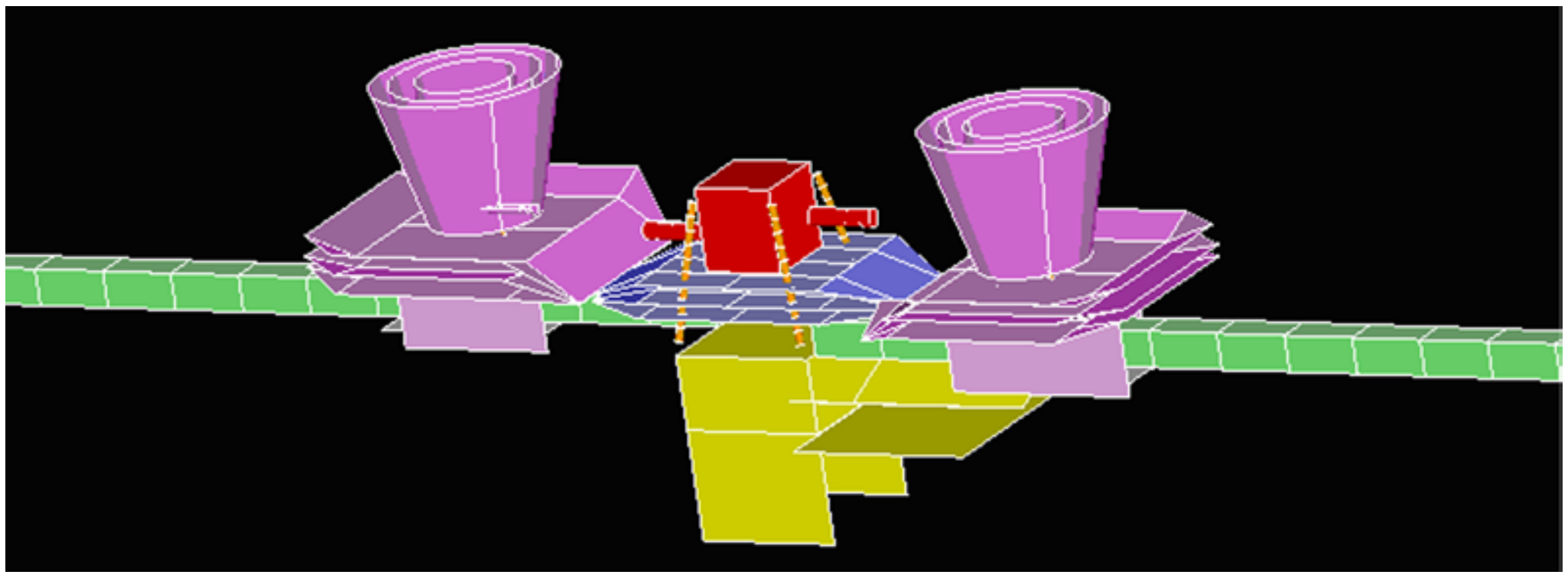

**Figure 7** – SPIRIT's shades and shields provide passive and active cooling, prevent stray thermal radiation from reaching the focal plane, use space economically, and provide access to the required field of regard. The radiators, positioned far below the centrally-located spacecraft, are the only deployable elements of the thermal-cryo system.

### 3.3. Detectors

The design attributes described in Section 3.2 will limit the undesirable radiation reaching SPIRIT's detectors to a small fraction of the celestial light intensity from the targeted region. Additionally, to meet the SPIRIT sensitivity requirement, the detector noise must contribute negligibly (<10%) to the output signal. Upper limits to the tolerable detector Noise Equivalent Power (NEP) were derived for SPIRIT from the accurate COBE absolute sky brightness measurements.[53] Each of the four SPIRIT wavelength bands has its own dedicated set of detector arrays. The detector NEP must not exceed 1.9, 1.1, 0.7, or 1.8 x $10^{-19}$ W $Hz^{-1/2}$ in the 25 – 50, 50 – 100, 100 – 200, and 200 – 400 μm wavelength bands, respectively.

We employ the spatial multiplexing technique being developed at GSFC under the NASA ROSES/APRA-funded "Wide-field Imaging Interferometry" study,[49, 50] in which adjacent detector pixels record Michelson interferometric fringes from contiguous regions of the sky to cover the desired field of view. In the SPIRIT C design, the detector pixels Nyquist sample the Airy disk of a 1 m telescope at the geometric mean wavelength in each of the four octave-wide wavelength bands. Accordingly, the detector array for the shortest wavelength channel has 14 x 14 pixels, and the arrays for each successively longer wavelength channel contain 7 x 7, 4 x 4, and 2 x 2 pixels. Additional experimentation and analysis are needed to determine if sparser-than-Nyquist sampling is preferable, in which case the pixel count could be slightly reduced. The modest number of detector pixels and the high observing efficiency achievable when detector arrays are used instead of observatory pointing to cover the field of view are important features of the SPIRIT design.

From the requirement that SPIRIT be capable of densely sampling the u-v plane and providing R = 3000 spectroscopy over a 1 arcminute FOV in 1 day we derived the optical modulation rate resulting from a scan of the delay line and the corresponding detector time constant. The detector time constant must not exceed 185 μs.

Transition-Edge Sensor (TES) bolometers cooled to 30 mK, with SQUID amplifiers for readout, are theoretically capable of satisfying the SPIRIT detector noise and speed requirements, and recent lab work has yielded promising results.[54 - 56]

### 3.4. Metrology

To minimize instrumental degradation of the fringe visibility, light from both arms of a Michelson interferometer must arrive at the combination plane with a large fractional beam overlap and a small angular wavefront misalignment (specifically, misalignment by no more than a small fraction of a wavelength across the pupil plane). We required a visibility of at least 0.94 on an unresolved reference source prior to calibration (by definition, the calibrated visibility is 1.0 in this case). In other words, the visibility loss due to all instrument imperfections, including optical surface imperfections, totals less than 6%.

An internal metrology system provides the information needed to maintain pointing and optical system alignment and stabilize the optical path length, and it provides knowledge of the baseline length and the time-varying optical path difference between the two arms of the interferometer during delay line scans. To accomplish these objectives, the metrology system directly measures certain dimensions of the opto-mechanical hardware, including the separations and relative orientations of mirrors in the light collecting telescopes and the beam combiner, and the separations between path length controlling mirrors inside the beam combiner, as shown schematically in Figure 6.

Guide stars are used to orient the spacecraft in absolute coordinates, and near-IR point sources in the science field of view serve as phase references for the optical path external to the instrument.

The metrology and control tolerances for SPIRIT are greatly relaxed relative to those for interferometers designed to operate at shorter wavelengths or null starlight, such as the Space Interferometry Mission (SIM)[57] and TPF-I/Darwin.[33,34] We assigned visibility loss tolerances to various instrument subsystems and derived subsystem requirements and descriptions of the metrology hardware components (path-length sensor, angle sensor, zero-path-difference sensor, laser ranging, and tip/tilt sensors) depicted in Figure 6. Some of the SPIRIT hardware is common to SIM, and we concluded that a modest investment in metrology technology development can lead to significantly lower total cost if advantage is taken of the more easily satisfied SPIRIT requirements.

### 3.5. Mechanisms

The mechanisms in SPIRIT are shown in Figure 6. Beam steering and path length control mechanisms provide a stable optical bench. At each telescope, a focusing mechanism adjusts the position of the secondary mirror, tip and tilt adjustments enable telescope pointing, and a trolley mechanism enables telescope movement to an arbitrary position along the boom structure to adjust the baseline length. A shutter mechanism equipped with an array of calibration lamps is located after the first optic in the beam combiner and can be used to calibrate the detector performance or operate the interferometer in “single-dish” mode to obtain low spatial frequency information. The optical delay line (ODL) scan mechanism provides linear motion and operates almost continuously throughout the mission. Devices similar to many of those required for SPIRIT, including the ODL scan mechanism, have been used successfully in space under similar operating conditions (temperature, rate of motion, frequency of usage, and lifetime).

Many of the mechanisms operate at 4 K. The parasitic heat loads from these cryogenic mechanisms will be limited so as not to exceed the capacity of the cryocoolers to remove heat from the system. The actuators in these mechanisms will take advantage of the low-temperature environment and rely on superconductivity to keep Ohmic losses to a minimum.

### 3.6. Deployment and operation

The preferred location for SPIRIT is the Sun-Earth Lagrange point L2, which offers deep radiative cooling and minimum interference from the Earth and Moon, while requiring only a single-sided sunshield. Delta IV and Atlas V launch vehicles were considered, and the latter was chosen due to its larger fairing dimensions and adequate lift capacity. The fairing volume limits the boom length and therefore the achievable angular resolution of the interferometer. Figure 8 shows SPIRIT in its stowed and fully-deployed configurations.

We developed a detailed description of the mechanical system, with a parts and mass properties inventory, and conducted structural analyses of SPIRIT in the launch and deployed configurations. Our structural analysis covered both dynamics and thermal distortion.

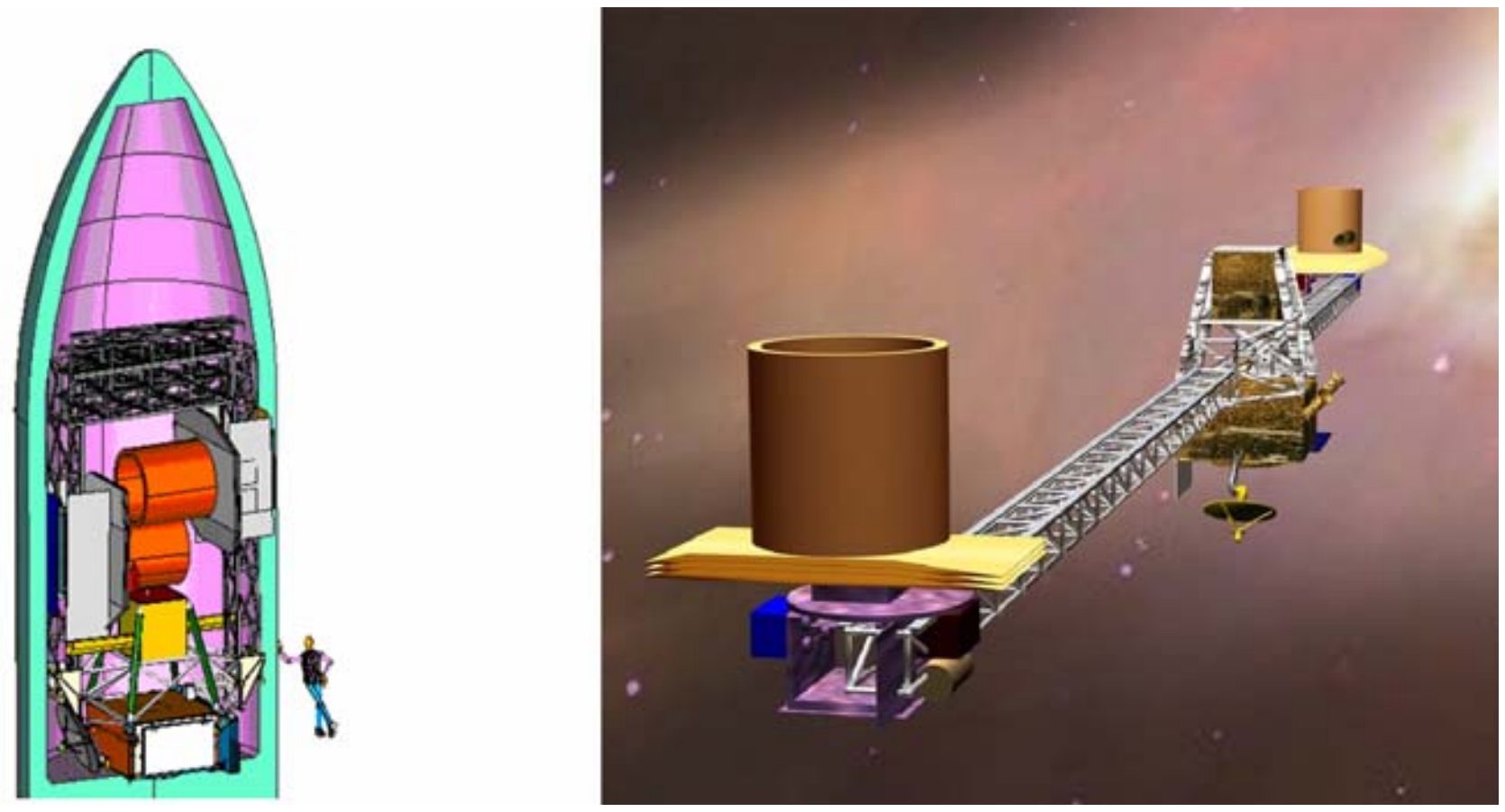

**Figure 8** – SPIRIT and its expendable launch support structure, when stowed for launch (left), are 8.7 m tall and were designed to fit into an Atlas V 5-medium fairing. The deployed observatory is shown in an artist’s concept (right).

The major post-launch mission activities are:

- Deployment and spacecraft checkout phase (first week) – deploy solar arrays, acquire Sun, deploy high-gain antenna, check out spacecraft subsystems (e.g., communications, attitude control, thrusters), deploy radiator, boom structure, and telescope transport trolleys;
- Instrument warm checkout and cool-down phase (weeks 2 – 5) – activate cryocoolers and check out each subsystem as soon as its temperature permits;
- Instrument cold checkout and science commissioning phase (weeks 6 – 12) – align and calibrate the instrument with real-time ground contact, and verify performance.

Four or more course correction maneuvers are interspersed with these activities prior to L2 insertion in the 15th week after launch. Normal operations begin after commissioning and will last for 3 to 5 years, interrupted about 4 times a year for station-keeping maneuvers. At the end of the mission, the spacecraft will be decommissioned by maneuvering it away from L2.

SPIRIT will follow a large-amplitude Lissajous orbit around Sun-Earth L2. The instantaneous field of regard is limited by the size of the sun shades to a +/-20 deg cone around the anti-Sun direction. Over the course of a year, SPIRIT will have a 40-degree wide viewing zone around the ecliptic plane. The science targets available in that zone (Figure 9) will enable us to fulfill the science objectives of the mission.

An observation sequence comprises a slew to the target field, target acquisition (including lock on angle and zero path difference tracking), and science data acquisition (scan delay line and rotate the observatory all the time; calibrate detectors; adjust telescope positions to sample a new baseline length after every half-rotation at specified time intervals until all of the desired baselines are measured). A typical observation will take about 28.2 hours, with 2 hours allocated for slew and settle, one hour for trolley movements, 24 hours for collecting science data, and 1.2 hours for data downlink. The observing efficiency is 85% in this case, and will be even higher for deep exposures, which could last up to ~$10^6$ seconds (i.e., 11 days, instead of the typical 1-day exposure time). SPIRIT will yield approximately 400 Gbits of data per day.

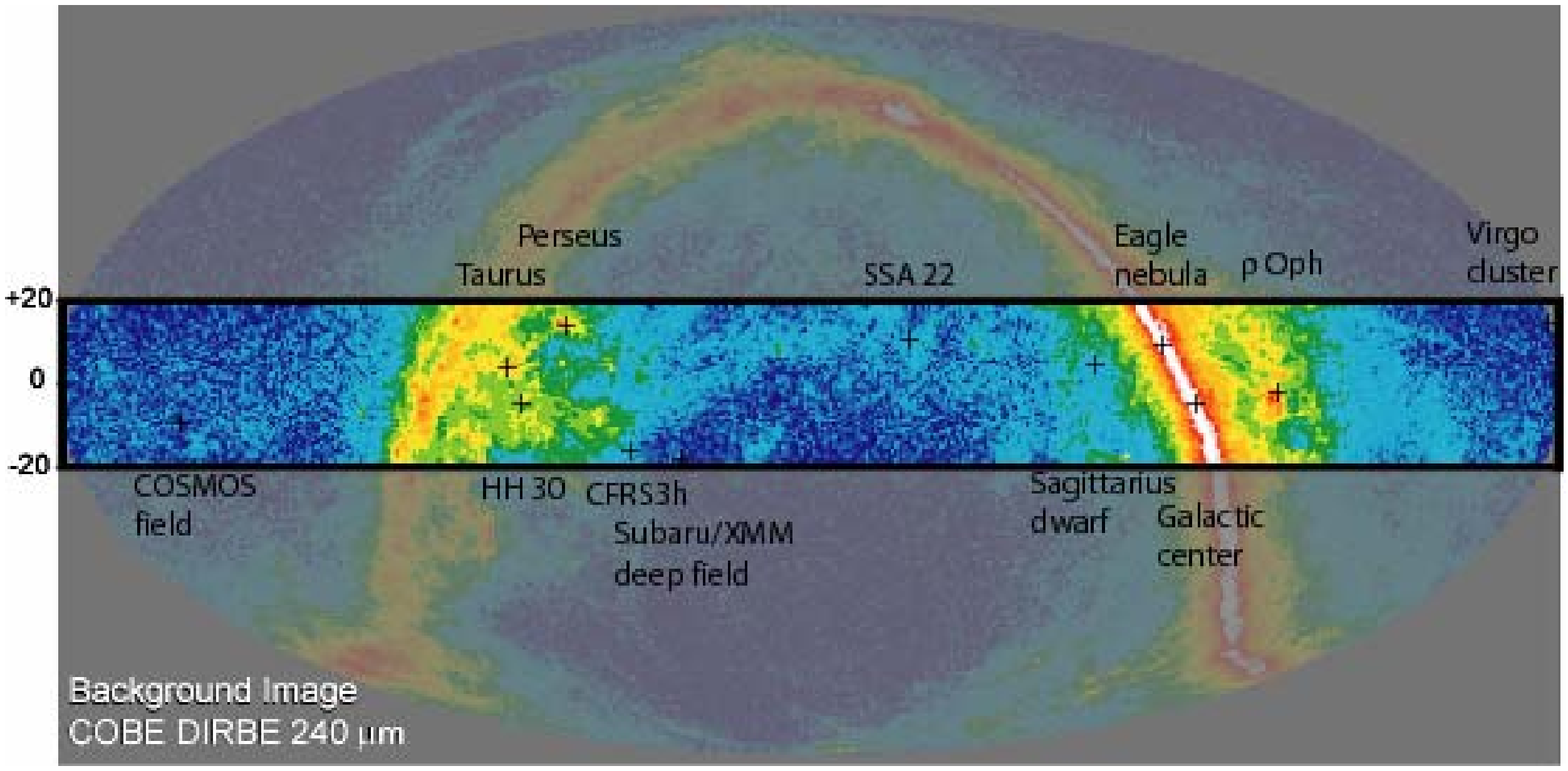

**Figure 9** – SPIRIT will have a view of the sky that permits access to all of the relevant object types, from Galactic star forming regions (e.g., Taurus and ρ Oph) to extragalactic "deep fields" (e.g., the COSMOS field and SSA 22). The titles in this figure correspond to the + symbols which lie above or below them and indicate the object positions.

SPIRIT has only one mode of operation, the double Fourier mode, but has the flexibility to observe a wide variety of astronomical sources with optimal science productivity. Operational parameters will be selectable by the observer, who will specify the desired spectral resolution (affects delay line scan range), field of view (affects number of pixels to read and, to a small extent, the delay line scan range), sensitivity (affects total observation time), and angular response function, or "synthesized beam" shape (affects the time spent at each interferometric baseline). This flexibility comes at a small price, primarily impacting the flight software. Observations can be queue-scheduled, giving mission operators the opportunity to optimize the sequence of target visits, and the parameter settings for each observation can be uplinked months in advance.

### 3.7. Integration and test (I&T) plan

Several aspects of the recommended I&T program are worth noting here, as they indicate how risks can be reduced to tolerable levels at an acceptable cost. In particular, we advocate: (a) incorporation of lessons from major missions currently in the formulation phase, such as JWST and SIM; (b) early development of a SPIRIT I&T Plan, and early consideration of test facility limitations, including their physical dimensions and geographic locations, which can affect the design and the costs and risks associated with shipping and handling flight hardware; (c) utilization of existing test facilities; (d) a "protoflight" approach which seamlessly flows from component technology validation to engineering test units, and ultimately to a flight-qualified, fully-integrated instrument; and (e) to the extent possible, the light collecting telescopes and the beam combining instrument module should be developed and tested in parallel, ground support equipment (GSE) should be reused, and facility use should be optimally scheduled. The instrument module is sufficiently compact that it can be performance tested independently of the light collecting telescopes, as shown in Figure 10. Development via a protoflight approach avoids unnecessary hardware replication and test repetition, leading to a more compact schedule and lower cost. The SPIRIT I&T plan outlines all the steps believed to be necessary to integrate, characterize, calibrate, and test the complete instrument functionally, for performance, and environmentally, using existing facilities.

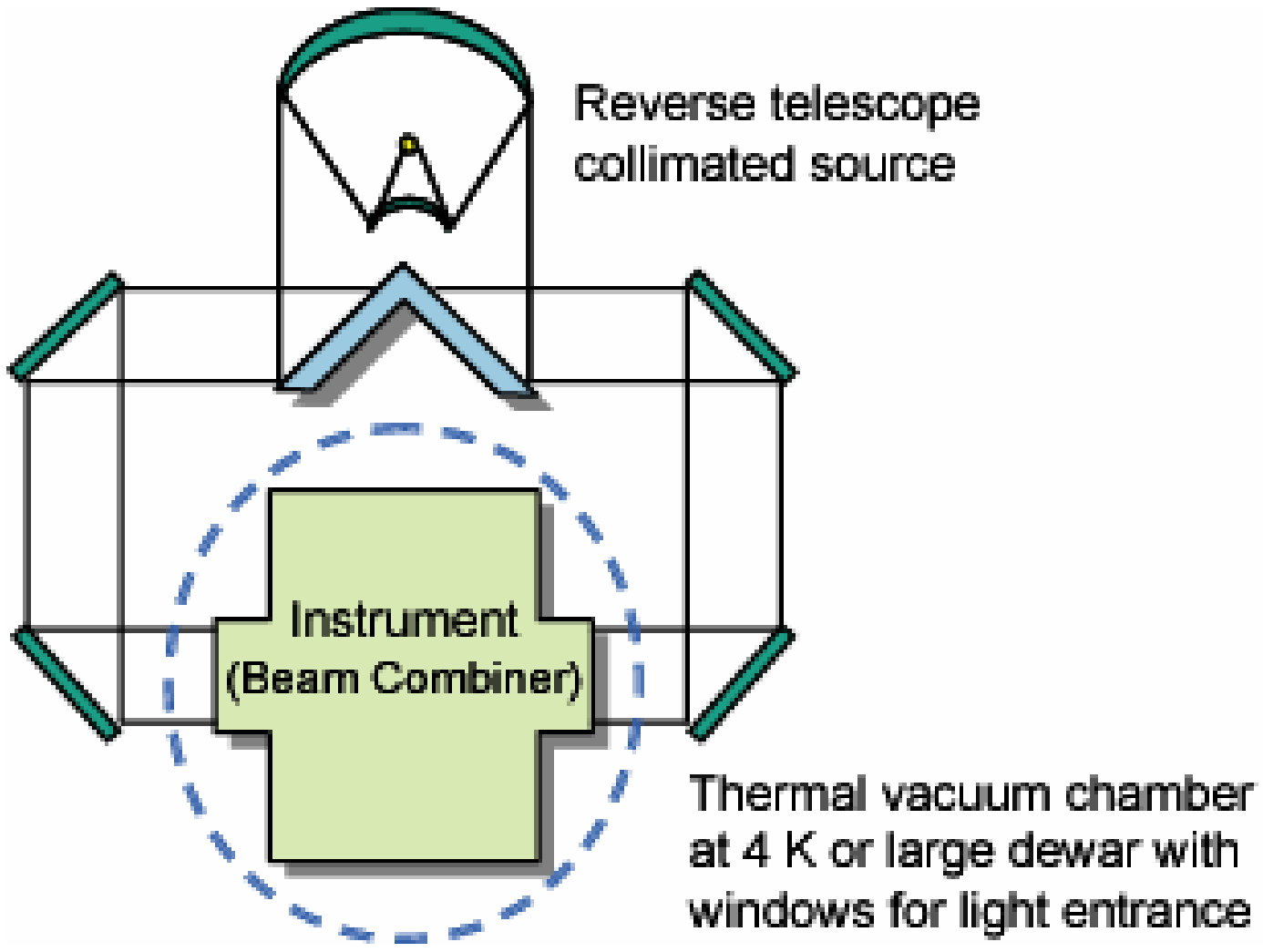

**Figure 10** – The beam combining instrument fits into existing thermal vacuum chambers and can be tested independently of the light collecting telescopes. The telescopes simply collect light and steer a collimated beam toward the beam combiner, so they can be tested individually in a similar manner.

## 4. MEASUREMENT CAPABILITIES COMPARED WITH THOSE OF OTHER FACILITIES

Ultimately, SPIRIT's ability to solve otherwise unsolvable fundamental scientific problems, rather than its unique measurement capabilities, make the mission compelling, but a straight comparison of measurement capabilities can also be informative. In Figures 11 and 12, respectively, we show the angular resolution and sensitivity of SPIRIT as a function of wavelength, and compare SPIRIT with current, near-term, and "vision" missions for the far-IR, and with JWST, which operates at shorter wavelengths, and ALMA, which operates at longer wavelengths.

Figure 11 shows that SPIRIT will bridge a gap in imaging capability between the longest wavelengths accessible to JWST and the shortest wavelengths accessible to ALMA. The canyon that needs to be spanned is several orders of magnitude deep and stretches across four octaves in wavelength.

SPIRIT will observe many astronomical objects where they emit the bulk of their radiation, and with angular resolution comparable to that of JWST. According to Wien's Law, thermal radiation sources whose temperatures range from 7 to 116 K glow most brightly in the SPIRIT wavelength range between 25 and 400 μm. Only the darkest interiors of starless molecular clouds could be colder than 7 K and have their peak emission at submillimeter wavelengths, and many of the interesting objects warmer than 116 K are concealed by dust and unobservable at UV and visible wavelengths.

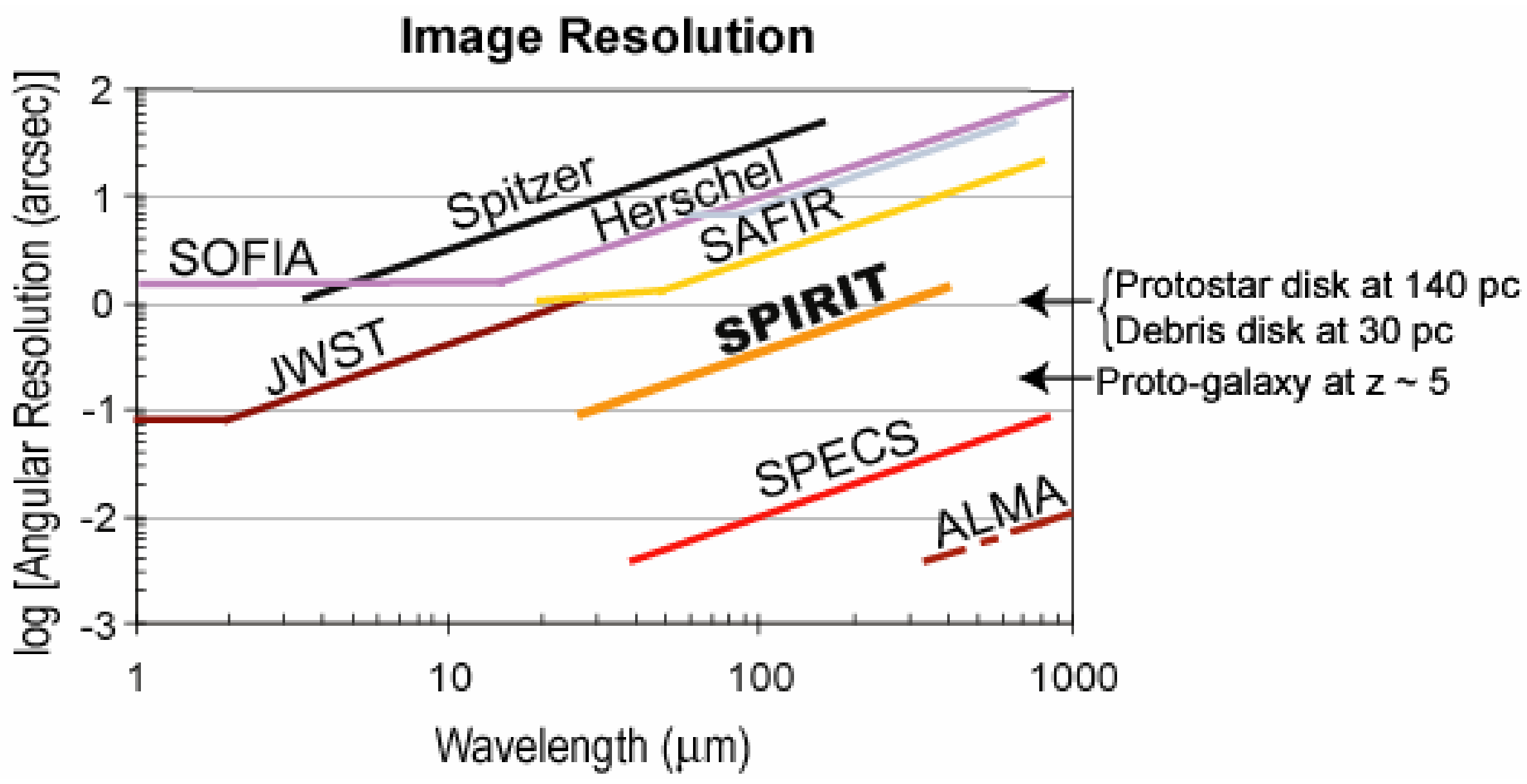

**Figure 11** – Angular resolution available with current, next-generation, and future infrared and submillimeter observatories. SPIRIT will provide one hundred times better angular resolution than the Spitzer Space Telescope. The resolution is comparable to that of JWST, but at ten times longer wavelengths.

With JWST, SPIRIT, and ALMA working in tandem, the astronomical community will have an extremely powerful arsenal of observing tools. JWST will follow the history of starlight back to its first appearance in a young universe and study the assembly of galaxies. ALMA will provide 0.1" images of submillimeter galaxies at redshift $z \sim 3$, measure their redshifts and characterize their reservoirs of star-forming gas. SPIRIT will make the complementary observations of the gas and dust, kinematics and morphology of galaxies at redshifts $z < 3$ needed to understand their subsequent development. Significant galaxy-building activity is thought to have taken place since the lookback time corresponding to $z = 3$. ALMA will image gas kinematics and study chemistry in protostars and protoplanetary disks. JWST will probe the low-mass end of the stellar initial mass function, study environmental effects on star formation, and use coronagraphy to investigate extrasolar planets and debris disks. SPIRIT will bring many new debris disk structures into the observable realm by providing 9x better angular resolution at 25 µm than JWST and better submillimeter dynamic range than ALMA. SPIRIT will penetrate dustier environments than those accessible to JWST and measure the spatial distributions of water and simple hydrides to complement ALMA's studies of gas kinematics and chemistry.

Figures 11 and 12 only partially illustrate the new discovery space that SPIRIT will open to exploration. To fully appreciate the enhancement in measurement capability, one must also recognize that, although SPIRIT's sensitivity will be comparable to that of the Spitzer Infrared Spectrograph (IRS),[45] SPIRIT provides this sensitivity with $R = 3000$ spectral resolution in the spectral range $25 < \lambda < 400$ µm, whereas the Spitzer IRS, the only spectroscopic instrument on the telescope, provides $R = 600$ in the wavelength range $10 < \lambda < 37$ µm and lower spectral resolution in the interval $5.3 < \lambda < 10$ µm. Spectral dilution at $R = 3000$ is not severe, and even relatively weak lines will be detectable with SPIRIT. Indeed, at $R = 3000$, SPIRIT will be able to measure the kinematics of merging protogalactic objects at high redshifts. Many of the spectral lines accessible to SPIRIT are astrophysically significant as probes of physical conditions or diagnostic tools. For example, the rotational and rovibrational lines of $H_2O$, CO and small hydrides probe temperatures, densities, and isotope abundances in molecular clouds and protostellar disks; the far-UV field strength, temperature and density in "photon-dominated" regions can be measured with observations of the fine-structure lines of $C^+$ at 158 µm, the neutral oxygen lines at 63 and 145 µm, and the neutral carbon lines at 370 and 609 µm.; and the [Ne V] (14.24 µm)/[Ne III] (15 µm) line ratio measures the ionization parameter in Active Galactic Nuclei (AGN).

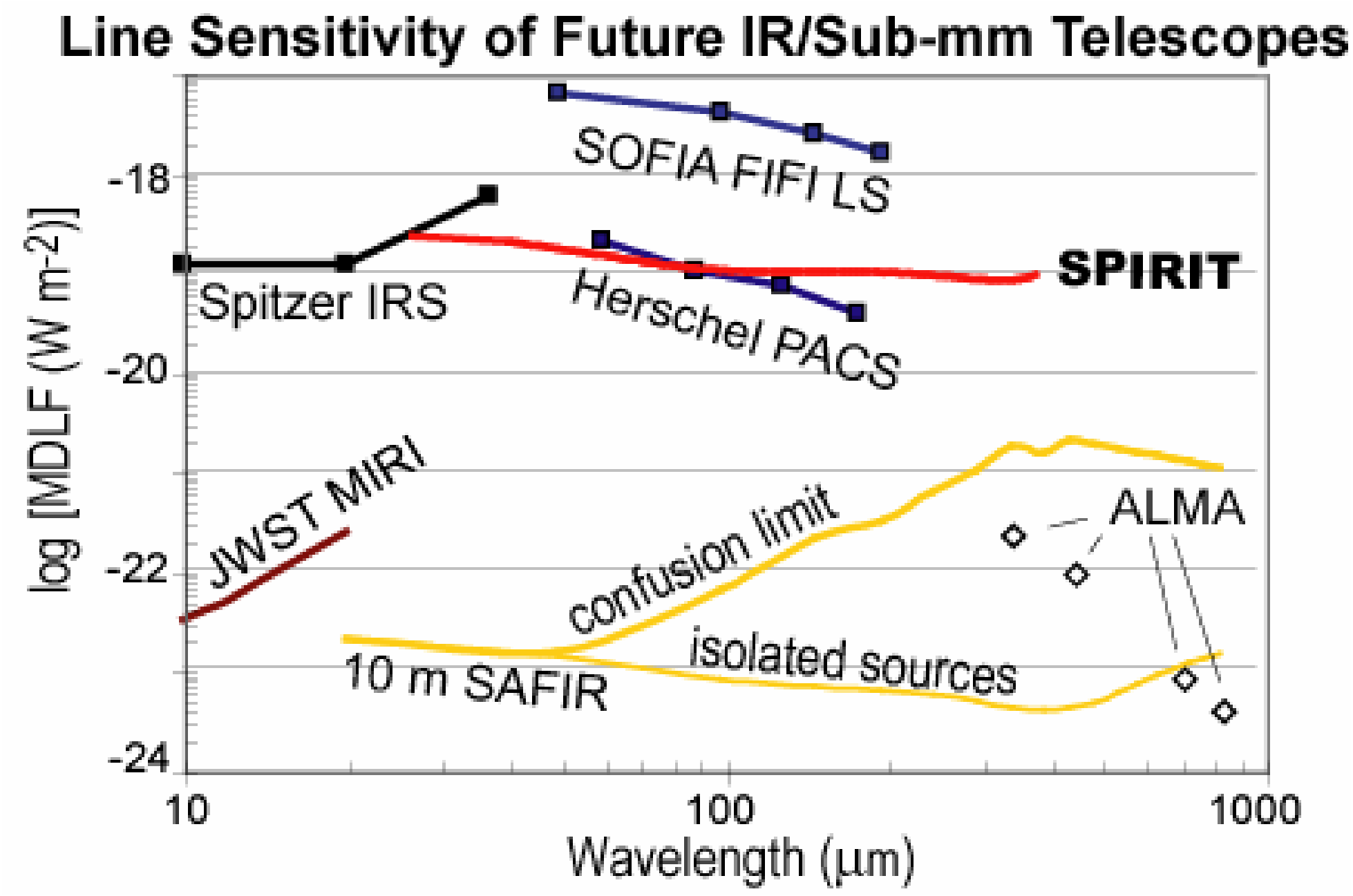

**Figure 12** – Minimum Detectable Line Flux (MDLF) available with current, next-generation, and future infrared and submillimeter observatories. SPIRIT will approximately match the Spitzer IRS sensitivity, but at higher spectral resolution and much higher angular resolution. SPIRIT has less light collecting area than SOFIA and Herschel, but SPIRIT's modest aperture area is compensated by cryogenically cooled optics. Quoted values are for a $10^5$ sec exposure, except $10^4$ sec for SOFIA.

## 5. SUMMARY

The Space Infrared Interferometric Telescope (SPIRIT) was selected for study by NASA as a candidate Origins Probe mission. SPIRIT is a two-telescope Michelson interferometer operating over a nominal wavelength range 25 to 400 μm and offering a powerful combination of spectroscopy ($\lambda/\Delta\lambda \sim 3000$) and sub-arcsecond angular resolution imaging in a single instrument. With angular resolution two orders of magnitude better than that of the Spitzer Space Telescope, and with comparable sensitivity, SPIRIT will enable us to learn how planetary systems form in protostellar disks, how they acquire their chemical structure, and how they evolve, and it will enable us to learn how galaxies formed and evolved over time. The key enabling technologies for SPIRIT are far-infrared detectors, cryocoolers, and cryogenic mechanisms.

## ACKNOWLEDGMENTS

Goddard's senior management made a generous resource commitment to the SPIRIT study, without which most of the results described here could not have been achieved. We are very grateful for their support. Special thanks to the SPIRIT Advisory Review Panel – Gary Melnick (SAO), Chair, Dave Miller (MIT), Harvey Moseley (GSFC), Gene Serabyn (JPL), Mike Shao (JPL), Wes Traub (SAO), Steve Unwin (JPL), and Ned Wright (UCLA) – for their expert advice. Four industry partners conducted parallel studies on selected topics and contributed their results, enhancing the success of the overall SPIRIT study. In particular, we thank Dave Fischer, Dave Glaister, Paul Hendershott, Vince Kotsubo, Jim Leitch, Jennifer Lock and Charley Noecker of Ball Aerospace, Tracey Espero, Ed Friedman and Mike Kaplan of Boeing, John Miles, Ted Nast, Jeff Olson, Dominick Tenerelli and Bob Woodruff of Lockheed-Martin, and Chuck Lillie and Ron Polidan of Northrop-Grumman. J. F. acknowledges the support of the Naval Research Laboratory 6.1 base funding. We thank our anonymous referees for their many helpful recommendations. NASA sponsored the SPIRIT study under the Origins Science Mission Concept study program.